\documentstyle[seceq,epsf]{ptptex}
\newcommand{\be}{\begin{equation}}
\newcommand{\ee}{\end{equation}}
\newcommand{\bea}{\begin{eqnarray}}
\newcommand{\eea}{\end{eqnarray}}
\newcommand{\beb}{\begin{eqnarray*}}
\newcommand{\eeb}{\end{eqnarray*}}

\newcommand{\up}{{\uparrow}}
\newcommand{\down}{{\downarrow}}
\newcommand{\SrCu}{SrCu$_2$(BO$_3$)$_2$}
\title{Magnetization process from Chern-Simons theory and its application to \SrCu}

\author{
Thierry {\sc Jolic\oe ur}\footnote{e-mail~:
Thierry.Jolicoeur@lpmc.ens.fr}, Gr\'egoire {\sc Misguich$^{**}$}
and Steven M. {\sc Girvin$^{***}$}}

\inst{$^*$Laboratoire de Physique de la Mati\`ere Condens\'ee,
Ecole Normale Sup\'erieure, 75231 Paris, France
\\
$^{**}$Service de Physique Th\'eorique, CEA Saclay, 91191
Gif-sur-Yvette, France
\\$^{***}$Sloane Physics Laboratory,
Yale University P.O. Box 208120 New Haven, CT 06520-8120, USA}

\recdate{\today }
\abst{In two-dimensional systems, it is possible transmute bosons
into fermions by use of a Chern-Simons gauge field. Such a mapping
is used to compute magnetization processes of two-dimensional
magnets. The calculation of the magnetization curve then involves
the structure of the Hofstadter problem for the lattice under
consideration. Certain features of the Hofstadter butterfly are
shown to imply the appearance of magnetization plateaus. While not
always successfull, this approach leads to interesting results
when applied to the 2D AF magnet \SrCu. }
\begin{document}
\maketitle
%%%%%%%%%%%%%%%%%%%%%%%%%%%%%%%%%%%%%%%%%%%%%%%%%%%%%%%%%%%%%%
\section{Introduction}

Among the accidents that may happen to the magnetization curve of
a magnet, the formation of a plateau has been observed already in
a variety of physical situations. In antiferromagnets with a
triangular lattice, a plateau at M/M$_{sat}$=1/3 has been
discovered in C$_6$Eu, CsCuCl$_3$ and RbFe(MoO$_4$)$_2$. On the
theoretical side, there is such a plateau at M/M$_{sat}$=1/3 for
the two-dimensional Heisenberg antiferromagnet on the triangular
lattice\cite{theo1}. This plateau has a semiclassical
explanation~: it is due to the stabilization of a collinear
$\up\up\down$ state. In a similar vein, there is a plateau at
M/M$_{sat}$=1/2 in a model with a four-spin exchange which is due
to stability of a $\up\up\up\down$ state\cite{theo2}.

Plateau formation also happens in a series of one-dimensional
systems, both experimentally and theoretically. In some cases,
they can be explained by a limit in which one has a dimerized
system, i.e. this happens for the alternating S=1 spin chain.
When, say, odd bonds are much stronger than even bonds then we
have a set of essentially decoupled dimers and if we apply a
magnetic field upon such a set of dimers there is a two-step
magnetization process. If we now add perturbatively a coupling
between the dimers, the plateau will of course survive at least
for a finite range of parameters. There are also plateaus that can
be understood in terms of a band filling picture. This happens for
the alternating ferro-antiferro S=1/2 spin chain. In the XY limit
it is a system of free fermions by the Jordan-Wigner mapping.
These fermions have a 1D band structure with gaps due to the
modulation of hopping from the alternating F-AF pattern. These
gaps then naturally corresponds to magnetization plateaus.

If we consider now again  2D systems, then similarly some of the
plateaus that are known may be explained via a dimerized
structure. It is tempting to speculate that some others may be due
to nontrivial band-filling effects for fermions living on the
lattice defining the magnet. The most intriguing 2D magnet with
plateaus in M(H) is, up to now, the compound\cite{Kageyama99}
\SrCu. It is a 2D antiferromagnet with localized  spins S=1/2, a
spin gap\cite{Miyahara1999} and its magnetization curve  has
plateaus\cite{Onizuka2000} M/M$_{sat}$=1/3, 1/4, 1/8. These
plateaus are observed by using fields as high as 57~T. There may
be even more plateaus waiting for us in this system. The magnetic
ions Cu$^{2+}$ have a peculiar lattice structure which is shown in
Fig.~1. This is the so-called Shastry-Sutherland lattice, a model
system that was invented theoretically many years
ago\cite{Shastry81}. This system is a square lattice with AF
exchange J' and there are additional bonds on some of diagonals
that we note J. It is easy to guess that when J is large enough
the system will be dimerized but in fact the Shastry-Sutherland
lattice has the remarkable peculiarity that the direct product of
local dimers on the J bonds is an {\it exact} eigenstate of the
Hamiltonian. This is true for all values of the couplings and due
to the fact that the coupling involves only triangles of spins.
This dimerized state is the ground state when J is large enough.
For J'  smaller  than $\sim 0.7~J$~\cite{Weihong99}, the system
has   dimer long-range   order and  for  larger J' it has
conventional N\'eel  long-range  AF order since when J=0 we find
the known square lattice AF magnet. There may be additional phases
inbetween like a plaquette singlet phase\cite{Koga2000} but they
are probably not realized in  \SrCu where  J'/J estimated to be
smaller than 0.65~\cite{Miyahara2000b,Knetter2000}. In fact,
susceptibility data as well as specific heat pointed to a value
J'/J=0.68 but taking into account three-dimensional couplings
leads to a revised estimate\cite{Miyahara2000b} J'/J=0.635.

\begin{figure}
\epsfxsize=6cm $$ \epsfbox{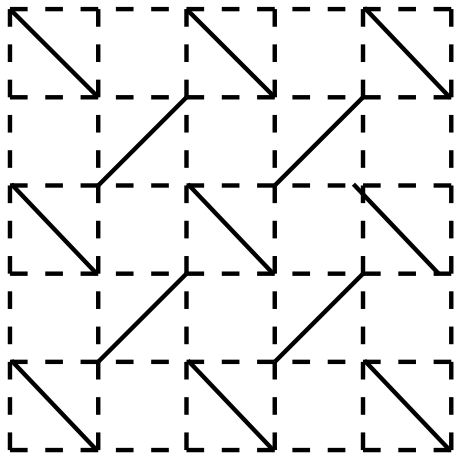}
 $$
 \caption{Shastry-Sutherland lattice.
Full (resp. dotted) lines denote the coupling $J$ ($J'$).}
\label{ShastryLattice}
\end{figure}

If we concentrate on the dimerized phase, then it naturally
explains the appearance of a spin gap. The first excited state
will be a triplet state formed by breaking a dimer and this
triplet will move on the lattice and form a band. In fact, due to
the peculiar triangular coupling which is responsible for the
exact eigenstate property, the motion of the triplet arises only
at high order in perturbation theory. As a consequence the
dispersion of the triplet band is very weak. This fact has been
confirmed by direct neutron inelastic
scattering\cite{KageyamaNeutrons}. Curiously, the two-triplet
states are on the contrary strong dispersions. The slow motion of
the elementary triplet has led to the hypothesis that they may
form some kind of crystal states when they are present at a finite
density which means finite magnetization. A first line of attack
has used an effective theory in which only the lowest-lying member
of the triplet is kept (since there is Zeeman splitting) and it is
treated as an effective hard-core boson. This has lead to the
prediction of the 1/3 plateau\cite{Momoi2000a} which has been then
discovered experimentally. In fact the effective bosonic particles
may display a variety of quantum phases\cite{Momoi2000b}. There is
also a related exactly solvable model with
plateaus\cite{Muller2000}. So far there is no microscopic probe of
the nature of plateau states.

If we want to investigate the possibility that band-filling
effects have something to do with the plateaus of \SrCu, we first
have to formulate the spin model in terms of fermions. This is
done in Section II. Then in section III, we explain how to derive
the magnetization curve in this mean-field approach. Results for
various lattices including the Shastry-Sutherland lattice are
given in section IV. Finally some conclusions and open problems
are discussed in section V.

\section{Chern-Simons statistics transmutation}

 The Hamiltonian for the AF Heisenberg
model on the Shastry-Sutherland lattice is~: \be H
=\sum_{\left\langle i,j\right\rangle} J_{ij} \quad {\vec
S}_{i}\cdot {\vec S}_{j} -B \sum_i S^z_i , \label{HamSpin} \ee
where   $\vec  S_i$  are  spin-1/2  operators,  the exchange
couplings $J_{ij}$ are equal to  J' when $i ,j$ are
nearest-neighbors on the square  lattice and equal to J   when $i
,j$  are  related by a diagonal  bond ($J, J' >  0$) and the
external  magnetic field $B$ is applied along the z-axis.  We then
use the standard mapping of the spin operators to hard-core boson
operators~: \bea H&=&H_{XY} + H_Z ,\\ H_{XY}&=&{1\over
2}\sum_{\left\langle i,j\right\rangle} J_{ij} \left( b^{\dag}_i
b_j + b^{\dag}_j b_i \right) ,\\ H_Z&=&\sum_{\left\langle
i,j\right\rangle} J^z_{ij} (n_i -1/2)(n_j -1/2)-B\sum_i (n_i
-1/2), \label{HamBose} \eea where $n_i\equiv S^z_i +1/2$ is the
occupation number of site $i$. The problem is now a lattice boson
system. To map it onto fermions, one has to generalize the
well-known Jordan-Wigner transformation to two space dimensions.
The corresponding recipe is well-known in the continuum limit. One
has to attach a flux tube onto each fermion and tune the flux to
exactly one flux quantum $\phi_0$. If we call $\varrho (r)$ the
fermion density, we have to require~: \be \nabla\times
{a}=\phi_0\varrho (r)\hat{z}. \label{constraint}\ee The vector
potential ${a} (r)$ can be computed if we take a density which is
a sum of delta functions~: \be
a(r_i)=\frac{\phi_0}{2\pi}\sum_{j\neq i}\nabla_i \arctan \left(
\frac{y_j-y_i}{x_j-x_i}\right). \ee There is now a nontrivial
phase~: \be \alpha_{ij}=\arctan \left(
\frac{y_j-y_i}{x_j-x_i}\right)\equiv\alpha (r_i-r_j), \ee which
maps fermionic wavefunctions onto bosonic wavefunctions through
the following unitary operator~: \be U=\exp\left[
i\sum_{i<j}\alpha_{ij}\right]. \ee Indeed if $\phi$ is bosonic
then $\Psi$ given by~: \be \Psi (r_1,\dots,r_N)=U  \phi
(r_1,\dots,r_N)\ee is a fermion wavefunction which has the same
eigenvalue as the bosonic problem provided one adds the gauge
field $a$ in the Hamiltonian. This is due to the fact that the
following identity is true~: \be Up_iU^{-1}=p_i -e\, a(r_i). \ee
Here $e$ appears because $\phi_0=2\pi /e$. The fake gauge field
$a$ has to obey Eq.(\ref{constraint}) and this can be enforced
through a very special Lagrangian which is permissible only in two
space dimensions~: \be L_{CS}=\int d^2 r\quad a_0
(\frac{e}{\phi_0}\varepsilon_{\alpha\beta}\partial_\alpha a_\beta
-e\varrho ) .\ee The time component of the gauge field $a_0$ is a
Lagrange multiplier. This is the so-called Chern-Simons
Lagrangian. It is at the heart of a very interesting mean-field
theory for the fractional quantum hall effect. For a detailed
exposition we refer to the book by Nagaosa\cite{NagaosaBook}.

If we want to perform the same kind of
trick\cite{Fradkin89,Ambjorn89} on the lattice Bose model, then
there is a part of the method which is straightforward. It is the
coupling of the gauge fields. These fields live naturally on the
bonds of the lattice so they enter only the expression of
$H_{XY}$~: \be H_{XY}={1\over 2}\sum_{\left\langle
i,j\right\rangle} J_{ij} \left( b^{\dag}_i b_j + b^{\dag}_j b_i
\right)\equiv {1\over 2}\sum_{\left\langle i,j\right\rangle}
J_{ij} \left( c^{\dag}_i {\rm e}^{ ia_{ij}} c_j + c^{\dag}_j {\rm
e}^{ -ia_{ij}} c_i \right) \ee The part $H_Z$ is unchanged since
only a phase relates bosons and fermions, hence $b^\dag b=c^\dag
c$. The relation between flux and density is not as simple as in
the continuum formulation. This is because gauge fields are living
on bonds while matter (fermions) is living on sites. The correct
prescription is to attach the flux of a site $x$ onto only one
plaquette which touches the site, say the upper right plaquette on
a square lattice. This is discussed at length in the work of
Eliezer and Semenoff\cite{Eliezer92}. This is of no importance for
the results we discuss because we next perform a mean-field
approximation. Also the lattice version of the Chern-Simons
Lagrangian is quite complicated but is of no concern here. The
only thing that matter is that density is tied to flux by
$\Phi=2\pi\rho$(x) where $\rho$(x) is the fermion density at site
x and $\Phi$ is the flux piercing the "corresponding" (say upper
right) plaquette {\it of the square lattice}. This mapping is
exact in the sense that it involves no approximations (even if it
is probably not rigorous~: the Eliezer-Semenoff construction has
been done for the square lattice only, so far). To use this
mapping one has to perform a mean-field
treatment\cite{Yang93,Lopez94} which is detailed in the next
section.

It is important to note that the fake magnetic field piercing the
plaquettes has nothing to do with the externally applied real
magnetic field. Indeed the applied field is a chemical potential
for the system of bosons as well as for the system of fermions.

\section{Mean-field treatment of the magnetization process}

To compute physical quantities with the lattice fermion model
coupled to the Chern-Simons gauge field, we perform the so-called
"average flux approximation". It amounts to replace the gauge
field a static uniform gauge field whose average flux is defined
by the average particle number on any site~:
\be
\Phi\equiv2\pi \langle \varrho \rangle . \ee In addition we have
to treat the Ising term $H_Z$ in Eq.(\ref{HamSpin}). This is done
by the simplest mean-field decoupling. As a consequence the
fermionic problem is now reduced to a one-body problem. This is
the problem of fermions on a lattice pierced by a uniform flux.
This very old problem has been investigated by many people
including Wannier, Azbel and Hofstadter\cite{Hofstadter76}.

We first discuss the procedure for the square lattice. The
tight-binding Hamiltonian on the square lattice with
nearest-neighbor hopping only has a single band in the absence of
flux. If the flux is rational, $\Phi/2\pi =p/q$ then this single
band disintegrates into $q$ magnetic subbands. This leads to an
infinitely complicated pattern of bands as a function of the flux,
the celebrated Hofstadter butterfly\cite{Hofstadter76}. If we now
consider a state of given magnetization for the fermion system,
then this leads to a given flux since we have~:
\be
\frac{\Phi}{2\pi}   = \langle \varrho \rangle
            = (\langle S^z \rangle +{1\over 2})
            = M+{1\over 2}.
\ee For example if $M=0$ then we have flux $\Phi=\pi$ and the
system is half-filled. This is the case $q=2$ hence there are two
subbands and we obtain the ground state by filling the lower band
with the available fermions. In the case of the square lattice
these two subbands do touch (but without overlapping). If we
consider the state for which $M/M_{sat}$=1/3 then we have $M$=1/6
and $\Phi/2\pi = 2/3$ there are three subbands and since $\langle
\varrho \rangle$ is also 2/3 the ground state is then obtained by
filling the lowest two subbands. By adding the energies of the
occupied states we get the energy of the fermion system. If we
perform such a calculation, we do not obtain directly the energy
as a function of the applied field but we obtain in fact the
Legendre transform of the energy as a function of the
magnetization $E(M)-BM$ (see Eq.(\ref{HamBose})). So $M$ as a
function of $B$ can be obtained by minimizing this quantity. A
plateau in the magnetization curve is a point on the $E(M)$ curve
at which there is a discontinuity in the {\it slope}. It may also
happen that a given slope (a $B$ value) is realized at {\it two}
points, this leads to a metamagnetic jump in the magnetization
curve. All these phenomena in fact do happen in the Chern-Simons
mean-field theory for antiferromagnets.

It is straightforward to get $E(M)$ numerically because this is
only a one-body problem. One has to diagonalize a tight-binding
problem with flux and fill the bands accordingly. The Legendre
transform is then also performed numerically. We discuss the
results of this approach\cite{MJG2001} for various lattices in the
next section.

\section{Results for various lattices}

\begin{figure}
\epsfxsize=6cm $$ \epsfbox{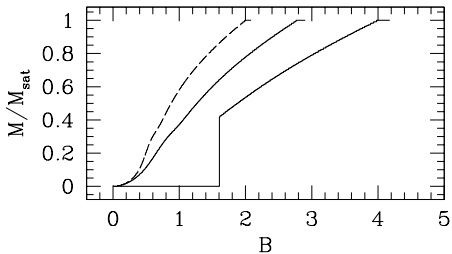}
 $$
 \caption{Magnetization curve for the square lattice model with
 use of a self-consistent mean-field with two-sublattices.
From left to right:  $J^{z}=0$,  0.39 and 1.  N\'eel long-range
order  appears  at $J^{z}\simeq0.39$  (Lopez  {\it et   al.}), for
higher values an Ising gap opens at $M=0$.} \label{square}
\end{figure}

We first discuss the case of the square lattice. From the
Hofstadter paper, it is clear that there is always a huge
fermionic gap just above the highest occupied state for all
magnetizations. The highest occupied state never jumps across the
gap even if it follows a fractal curve with the flux . As a
consequence the energy as a function of flux (hence magnetization)
is {\it smooth}. The magnetization curve of the square lattice is
then featureless. This is the left curve of Fig.~\ref{magcurves}.
Since the density of fermions is uniform, this is of course not a
valid description of a magnet with N\'eel order. This can be
improved by using a mean-field generalized to allow for two
sublattices. The difference of fermion densities is then found by
imposing self-consistency. Such an approach was pioneered by
Lopez, Rojo and Fradkin\cite{Lopez94}. The results are given in
Fig.~\ref{square}. As a function of Ising-type anisotropy, one
sees the appearance of N\'eel order. The correct value of the
transition to N\'eel order $J^z=1$ is not correctly recovered but
the phase structure is correct.

\begin{figure}
\epsfxsize=6cm $$ \epsfbox{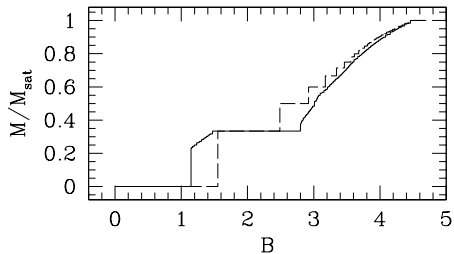}
 $$
 \caption{Magnetization curve of the triangular lattice.
Full line:  mean-field with three-sublattices. Dashed line:
uniform mean-field.} \label{triangle}
\end{figure}

In the case of the triangular lattice, we formulate the
Chern-Simons gauge theory by deciding that there is flux $\Phi/2$
piercing each triangular plaquette. If we use a uniform mean-field
then the resulting magnetization curve is strongly irregular~: it
has many plateaus as can be seen in Fig.~\ref{triangle}. There is
also a prominent zero-field gap~\cite{Yang93}. It is widely
believed that, instead, this spin system is gapless and its
magnetization   process shows only a single plateau at
$M/M_{sat}=1/3$.  To improve these results, we allow the
mean-field to have a three-sublattice structure: one introduces
three fermion densities $n_A,  n_B, n_C$ and numerically search
for a self-consistent solution where  the flux $\phi_\alpha$
matches the density $n_\alpha$ on each sublattice. For  $M=0$ the
self-consistent solution remains uniform. However, for non-zero
magnetization the translation symmetry  is broken. This
non-uniform mean-field solution  leads to     a magnetization
curve   shown Fig.~\ref{triangle}  which is much closer to    the
truth albeit the zero-field ground  state remains unrealistic. The
1/3 plateau has a semiclassical origin~: it derives from the $uud$
state that appears in the Ising limit $J^z\gg J$.

\begin{figure}
\epsfxsize=6cm $$ \epsfbox{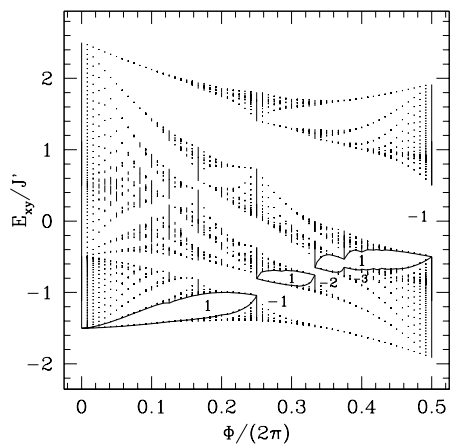}
 $$
 \caption{The Hofstadter butterfly for the Shastry-Sutherland lattice.
 Some hall conductances are given in the corresponding gaps. The lower thick
 line is the highest occupied state.}
\label{Hofstadt}
\end{figure}

We now discuss the Shastry-Sutherland lattice. The Hofstadter
butterfly for $J=J'$ is given in Fig.~\ref{Hofstadt}. It is clear
that there are jumps  of the Fermi energy as a function of $M$.
This leads  to discontinuity of the slope of the function  $E(M)$
and thus these jumps corresponds  to plateaus  in the
magnetization curve. The appearance of the plateaus is coded into
the delicate structure of the Hofstadter butterfly. It is
interesting to note that when there is an integer number of filled
subbands, then one has a well-defined quantized Hall conductance.
There is a recipe to compute this Hall conductance in the case of
the square lattice\cite{TKNN}~: one has to solve a simple
Diophantine equation. We have computed some of the Hall
conductances for the Shastry-Sutherland lattice by continuity
arguments (as integers of topological origin, they can only be
changed
 by  gap closure). These conductances are given in
Fig.~\ref{Hofstadt}.

\begin{figure}
\epsfxsize=6cm $$ \epsfbox{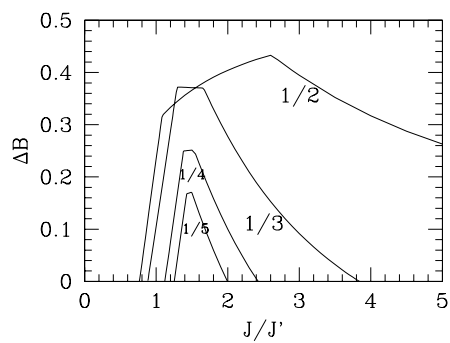}
 $$
 \caption{Regions of appearance of the magnetization plateaus as a function of J/J'
for $M/M_{\rm sat}=\frac{1}{2}, \frac{1}{3}, \frac{1}{4}$ and
$\frac{1}{5}$. Additional plateaux at fractions $\frac{1}{n}$ for
$n>5$ also exist in the vicinity of $J/J'=1.5$.} \label{plat}
\end{figure}

\begin{figure}
\epsfxsize=6cm $$ \epsfbox{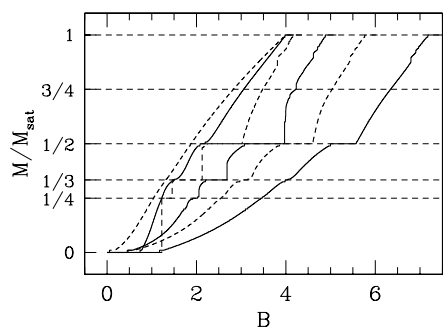}
 $$
 \caption{Magnetization curves for the Shastry-Sutherland lattice
 as a function of $J'/J$. From left to right $J=0$ (dashed) , 0.75
(full), 1.5(dashed), 2.5 (full), 3.5 (dashed) and  5 (full)
($J'=1$). } \label{magcurves}
\end{figure}

When the Fermi level follows smoothly the lower part of an "eye"
in the butterfly, we observe that the Hall conductance is +1. This
is due to the Streda formula. Indeed if we add one flux quantum to
the system we also add one fermion. If this fermion is going to
fill exactly one extra state at the bottom of the "eye" it means
that there should be exactly one extra state coming across the gap
from the upper states. This change of the number of states is
equal to the Hall conductance as shown by the Streda formula.
Changing the Hall conductance imply closing a gap and this is
exactly when there is a jump of the Fermi level. In this case we
see a plateau in the magnetization process. The domain of
stability of plateaus are given in Fig.~\ref{plat}.

\begin{figure}
\epsfxsize=6cm $$ \epsfbox{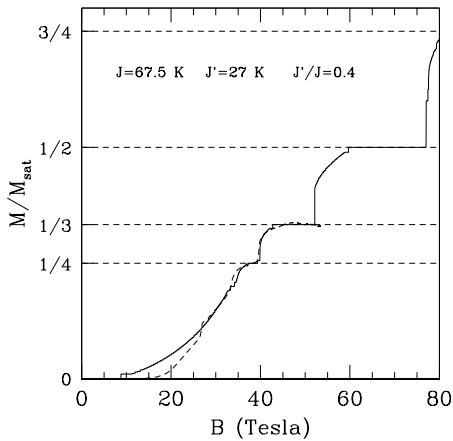}
 $$
 \caption{Best fit of the experimental magnetization curve of \SrCu.}

\label{fit}
\end{figure}

If we try to reproduce the  qualitative shape of the experimental
 curve for \SrCu, it is best to use  $J'= 29.5$K and $J=74$K,
 the  resulting fit is shown   in Fig.~\ref{fit}. These values do
not lead to a satisfactory spin gap~: 13K instead of 34K, but the
roundings close  to the plateaus are in good agreement with the
experiment. This approximation predicts prominent plateaus at 1/3
and 1/2  and no other plateaus till full saturation. There are
also small metamagnetic jumps just below saturation, the most
prominent one being at 3/4. We note that at 1/4 there is in fact
an avoided plateau~: the value of J'/J is just outside the range
of stability of the 1/4 plateau (see Fig.~\ref{plat}).

\section{Conclusions}

In the Chern-Simons mean-field approach, magnetization plateaus
appear a a manifestation of the integer quantum Hall effect for
fermions living on a lattice. These plateaus presumably survive
beyond mean-field. Indeed   Gaussian fluctuations\cite{Yang93} of
the gauge field are massive provided that the TKNN integer
$\sigma$ describing the  quantized Hall coefficient  of the
fermions on the frustrated lattice differs from the continuum
value of {\em unity}.  The plateaus are not suppressed since we
find $\sigma=-3$, $-2$ and $-1$ respectively for $\frac{1}{4}$,
$\frac{1}{3}$ and $\frac{1}{2}$. At $M/M_{sat}=0$ we have
$\sigma=-1$ (resp $0$) for $J/J'<\sqrt2$ (resp $J/J'>\sqrt2$).

This approach has at least some qualitative success in the case of
\SrCu. It is important to note that a consequence of these
non-trivial quantized Hall coefficients is that the spin state is
chiral and exhibits a `spin quantum Hall
effect'\cite{HaldaneandArovas}.

Finally we note that higher-field experiments should be able to
test the prediction for the magnetization curve of \SrCu.

\section*{Acknowledgements}
TJ would like to thank the organizers of the 16th
Nishinomiya-Yukawa memorial symposium. SMG is supported by grant
NSF DMR-0196503.

\end{document}